\documentclass[12pt,twoside,english,prb,preprint,showpacs]{revtex4}
\usepackage[T1]{fontenc}
\usepackage[latin1]{inputenc}
\usepackage{amsmath}
\usepackage{graphicx}
\usepackage{amssymb}

\makeatletter

\providecommand{\tabularnewline}{\\}

\usepackage{graphicx}

\usepackage{babel}
\makeatother

\begin{document}

\title{\emph{Ab initio} Wannier-function-based correlated calculations of
Born effective charges of crystalline Li$_{2}$O and LiCl}

\author{Priya Sony$^{*}$ and Alok Shukla}

\affiliation{Physics Department, Indian Institute of Technology, Powai, Mumbai
400076, INDIA}

\thanks{Present Address: Group of Atomistic Modeling and Design of Materials,
University of Leoben, Franz-Josef-Strasse 18, A-8700 Leoben, Austria}

\begin{abstract}
In this paper we have used our recently developed \emph{ab initio}
Wannier-function-based methodology to perform extensive Hartree-Fock
and correlated calculations on Li$_{2}$O and LiCl to compute their
Born effective charges. Results thus obtained are in very good agreement
with the experiments. In particular, for the case of Li$_{2}$O, we
resolve a controversy originating in the experiment of Osaka and Shindo
{[}Solid State Commun. 51 (1984) 421] who had predicted the effective
charge of Li ions to be in the range 0.58---0.61, a value much smaller
compared to its nominal value of unity, thereby, suggesting that the
bonding in the material could be partially covalent. We demonstrate
that effective charge computed by Osaka and Shindo is the Szigeti
charge, and once the Born charge is computed, it is in excellent agreement
with our computed value. Mulliken population analysis of Li$_{2}$O
also confirms ionic nature of the bonding in the substance. 
\end{abstract}

\pacs{77.22.-d, 71.10.-w, 71.15.-m}

\maketitle

\section{Introduction}

\emph{Ab initio} calculation of dielectric response properties of
materials are routinely performed using methods based upon density-functional
theory (DFT).\citep{dft-pol} Most of these calculations presently
are based upon the so-called 'modern theory of polarization' which
is based upon a Berry-phase (BP) interpretation of macroscopic polarization
of solids.\citep{king,resta} The practical implementation of the
BP approach within various versions of the DFT, such as the local-density
approximation (LDA), is straightforward because of their mean-field
nature. However, it is desirable to go beyond the mean-field level,
so as to compute the influence of electron correlation effects on
polarization properties such as the Born charge of crystals. Recently,
we have proposed an approach which allows computation of various polarization
properties of crystalline insulators within a many-body framework.\citep{priya-1,priya-2}
The approach utilizes a Wannier-function based real-space methodology,
coupled with a finite-field approach, to perform correlated calculations
using a Bethe-Goldstone-like many-body hierarchy. Its successful implementation
was demonstrated by performing \emph{ab initio} many-body calculations
of Born charges,\citep{priya-1} and optical dielectric constants,\citep{priya-2}
of various insulating crystals. We note that an \emph{ab initio} wave-functional-based
real-space correlated approach has recently been successfully demonstrated
by Hozoi \emph{et al.}\citet{hozoi}to compute the quasi-particle
band structure of crystalline MgO.

In this paper we apply our recently-developed approach,\citep{priya-1}
to perform \emph{ab initio} correlated calculations of Born charges
of crystalline Li$_{2}$O and LiCl. Li$_{2}$O is a technologically
important material with possible applications in thermonuclear reactors,\citep{osaka}
as also in solid-state batteries.\citep{farley} Based upon their
infrared reflectivity and Raman scattering based experiment on Li$_{2}$O,
Osaka and Shindo\citep{osaka} reported the value of effective charge
of Li ions to be in the range 0.58---0.61, a value much smaller compared
to the nominal value of unity expected in an ionic material. Therefore,
they speculated whether the nature of chemical bond in the substance
is partly covalent.\citep{osaka} In this work we present a careful
investigation of this subject, and resolve the controversy by showing
that the effective charge reported by Osaka and Shindo\citep{osaka}
was the Szigeti charge and not the Born charge. Once the Born charge
is computed from the Szigeti charge, its experimental value 0.95 is
in excellent agreement with our theoretical value of 0.91, and much
closer to the nominal value. We also present a plot of the Wannier
functions in the substance as also the Mulliken population analysis
to confirm the ionic bonding in Li$_{2}$O. As far as LiCl is concerned,
earlier we computed its Born charge at the Hartree-Fock level using
our Wannier function methodology,\citep{shukla-born} and found its
value to be significantly smaller as compared to the experiments.
Therefore, in this work, we perform correlated calculations of the
Born charge of LiCl, and, indeed, obtain much better agreement with
the experiments. 

Remainder of this paper is organized as follows. In section \ref{method}
we briefly describe the theoretical aspects of our Wannier-function-based
methodology. Next in section \ref{results} we present and discuss
the results of our calculations. Finally, in section \ref{conclusion}
we present our conclusions.

\section{Methodology}

\label{method} The Born effective charge tensor, $Z_{\alpha\beta}^{*}(i)$,
associated with the atoms of the $i$-th sublattice, is defined as\citep{born}\begin{equation}
Z_{\alpha\beta}^{*}(i)=Z_{i}+\left.(\Omega/e)\frac{\partial P_{\alpha}^{(el)}}{\partial u_{i\beta}}\right|_{{\textbf{E}}={\textbf{0}}},\label{zstar}\end{equation}
 where $Z_{i}$ is the charge associated with the nuclei (or the core)
of the sublattice, $\Omega$ is the volume of the unit cell, $e$
is magnitude of the electronic charge, and $P_{\alpha}^{(el)}$ is
the $\alpha$-th Cartesian component of the electronic part of the
macroscopic polarization induced as a result of the displacement of
the sublattice in the $\beta$-th Cartesian direction, $\Delta u_{i\beta}$.
For small $\Delta u_{i\beta}$, one assumes $\left.\frac{\partial P_{\alpha}^{(el)}}{\partial u_{i\beta}}\right|_{{\textbf{E}}={\textbf{0}}}=\frac{\Delta P_{\alpha}^{(el)}}{\Delta u_{i\beta}}$,
with $\Delta P_{\alpha}^{(el)}$ computed as the change in the electronic
polarization per unit cell due to the displacement $\Delta u_{i\beta}$.
For a given lattice configuration denoted by parameter $\lambda$,
we demonstrated that $P_{\alpha}^{(el)}(\lambda)$, for a many-electron
system, can be computed as the expectation value\citep{priya-1}\begin{equation}
P_{\alpha}^{(el)}(\lambda)=\frac{q_{e}}{N\Omega}\langle\Psi_{0}^{(\lambda)}|X_{\alpha}^{(el)}|\Psi_{0}^{(\lambda)}\rangle,\label{eq:corr}\end{equation}
$q_{e}$ ($=-e$) is the electronic charge, $N\:(\rightarrow\infty)$,
represents the total number of unit cells in the crystal, $X_{\alpha}^{(el)}=\sum_{k=1}^{N_{e}}x_{\alpha}^{(k)}$
is the $\alpha$-th component of the many-electron position operator
for the $N_{e}$ electrons of the crystal, and $|\Psi_{0}^{(\lambda)}\rangle$
represents the correlated ground-state wave-function of the infinite
solid, expressed in terms of Wannier functions. For one-electron models
such as Kohn-Sham theory, or the Hartree-Fock method, we showed\citep{priya-1}
that Eq. (\ref{eq:corr}) leads to expressions consistent with the
Berry-phase-based theory of King-Smith and Vanderbilt\citep{king},
and Resta\citep{resta}.

Calculating the expectation value of Eq. (\ref{eq:corr}) can be tedious
for a general correlated wave function. We avoid this by using the
finite-field approach whereby the expectation values of various components
of dipole operator can be obtained by performing both the HF and correlated
calculations in presence of a small electric field in that direction,
and then computing the first derivative of the computed total energy
with respect to the applied electric field.\citep{priya-1,priya-2}
This approach of computing expectation values derives its legitimacy
from the generalized Hellman-Feynman theorem, and is called {}``finite-field
method'' in quantum chemistry literature.\citep{mcweeny} Therefore,
the present set of HF and correlated calculations are distinct from
our previous ones\citep{shukla1,shukla2,shukla3,shukla4,shukla5}
in that the present calculations have been performed in presence of
an external electric field. For details related to our Wannier-function-based
HF methodology, in the presence of external electric fields, we refer
the reader to our earlier papers.\citep{priya-1,priya-2} 

Correlation contributions to the energy per unit cell $E_{corr}$,
even in the presence of external electric fields, were computed according
to the incremental scheme of Stoll\citep{stoll} \begin{eqnarray}
E_{corr} & = & \sum_{i}\epsilon_{i}+\frac{1}{2!}\sum_{i\neq j}\Delta\epsilon_{ij}+\nonumber \\
 &  & \frac{1}{3!}\sum_{i\neq j\neq k}\Delta\epsilon_{ijk}+\cdots,\label{eq:inc}\end{eqnarray}
where $\epsilon_{i},\:\Delta\epsilon_{ij},\:\Delta\epsilon_{ijk},\ldots$
etc. are respectively the one-, two- and three-body$,\ldots$ correlation
increments obtained by considering simultaneous virtual excitations
from one, two, or three occupied Wannier functions, and $i,\: j,\: k,\ldots$
label the Wannier functions involved.\citep{shukla4} Because of the
translational symmetry, one of these $n$ Wannier functions is required
to be in the reference unit cell, while the remaining ones can be
anywhere else in the solid. In our previous studies performed on ionic
systems, we demonstrated that the increments can be truncated to two-body
effects ($n=2)$, with the Wannier functions not being farther than
the third-nearest neighbors.\citep{shukla4,shukla5} We follow the
same truncation scheme here as well, with the method of correlation
calculation being the full-configuration-interaction (FCI) as in our
previous works, where other technical details related to the approach
are also described.\citep{priya-1,priya-2,shukla4,shukla5}

\section{Results and Discussion}

\label{results}

In this section, first we present the results of our calculations
of Born charges of Li$_{2}$O, followed by those of LiCl.

\subsection{Li$_{2}$O }

\label{sub:li2o}

The primitive cell of Li$_{2}$O consists of a three-atom basis (two
Li and one O), with each atom belonging to an fcc lattice. In our
calculations anion was placed at the origin of the primitive cell,
while the cations were located at the positions $(\pm a/4\pm a/4,\pm a/4)$.
This structure is usually referred to as the anti-fluorite structure,
to which other related materials such as Na$_{2}$O and K$_{2}$O
also belong. Based upon intuition, one would conclude that the bonding
in the substance will be of ionic type, with the valence $2s$ electrons
of the two lithium atoms of the formula unit being transferred to
the oxygen. Thus the two Li atoms will be in cationic state while
the oxygen will be in dianionic state. However, the argument for partial
covalence in the system goes as follows. It is well-known that the
free O$^{-2}$ ions do not exist in nature. Therefore, in lithium
oxide it will be stabilized only because of the crystal-field effects.
Still, one could argue, that O$^{-2}$ ion in solid state will be
very diffuse, leading to partial covalency. In light of these arguments,
it is interesting to examine the results of the experiments of Osaka
and Shindo\citep{osaka} performed on the single crystals of Li$_{2}$O,
where they measured the frequencies of fundamental optical and Raman
active modes using infrared and Raman spectroscopy. Using the measured
frequencies and several other parameters, the authors fitted their
data both to rigid-ion and shell models of lattice dynamics, to obtain
values for Li effective charges of 0.58 and 0.61,\citep{osaka} respectively.
Noting that the obtained effective charge is quite different from
the nominal value of unity, they argued that the result suggests that
Li$_{2}$O has partially covalent character.\citep{osaka} Next, we
argue, however, that the effective charge reported by Osaka and Shindo
is actually the Szigeti charge,\citep{szigeti} and not the Born charge.
The shell-model value of 0.61 of the effective charge of Li was computed
by Osaka and Shindo\citep{osaka} using the expression for the fluorite
structure derived by Axe\citep{axe}\begin{equation}
Z_{S}=[(\nu_{LO}^{2}-\nu_{TO}^{2})\mu\pi r_{0}^{3}]^{1/2}\frac{3\sqrt{\epsilon_{\infty}}}{\epsilon_{\infty+2}},\label{eq:szigeti}\end{equation}
where $\nu_{LO}/\nu_{TO}$ are the longitudinal/transverse optical
phonon frequencies, $r_{0}=\frac{a}{2}$, where $a$ is the lattice
constant, $\mu$ is the effective mass of the three ion system, while
$\epsilon_{\infty}$ is the high-frequency dielectric constant of
the material. The expression in Eq. \ref{eq:szigeti} is nothing but
the Szigeti charge\citep{szigeti} of the ion from which the Born
charge ($Z_{T}$) can be computed using the relation\citep{szigeti}\begin{equation}
Z_{T}=\frac{(\epsilon_{\infty}+2)}{3}Z_{S}\label{eq:born-szigeti}\end{equation}

If we use the values $\epsilon_{\infty}=2.68$ and $Z_{S}(\mbox{Li})=0.61$
obtained by Osaka and Shindo\citep{osaka} in Eq. \ref{eq:born-szigeti},
we obtain the experimental value of Born charge of Li ions in Li$_{2}$O
to be $0.95$ which is very close to the nominal value of unity. From
this, the experimental value of the Born charge of the O ions can
be deduced to be $1.90$. These values of the Born charges of Li and
O ions imply, in the most unequivocal manner, an ionic structure for
Li$_{2}$O consistent with the intuition, without any possibility
of covalency. 

Next, we report the results of our \emph{ab initio} correlated calculations
of Born charges of bulk Li$_{2}$O. These calculations were performed
using the lobe-type Gaussian basis functions which, as demonstrated
in the past, represent Cartesian Gaussian basis functions to a very
good accuracy.\citep{shukla1,shukla2,shukla3} As in our previous
HF study of the ground state properties of Li$_{2}$O,\citep{shukla3}
our calculations were initiated by a smaller basis set reported by
Dovesi \emph{et al.}\citep{dovesi} consisting of $(2s,1p)$ functions
for lithium, and $(4s,3p)$ functions for oxygen. However, we also
augmented our basis set by a single $d$-type exponent of 0.8, recommended
also by Dovesi \emph{et al.},\citep{dovesi} so as to check the influence
of higher angular momentum functions on our results. Henceforth, we
refer to these two basis sets as $(s,p)$ and $(s,p,d)$ basis sets,
respectively. For the purpose of the calculations, fcc geometry along
with the experimentally extrapolated $T=0$ lattice constant value
of $a=4.573$\AA\citep{farley} was assumed. For our finite-field
approach to Born charge described earlier, electric fields of strength
$\pm0.001$ atomic units (a.u.), along with an oxygen lattice displacement
$\Delta u=0.01a$, both in the $x-$direction, were used. This leads
to the determination of $Z_{xx}^{*}$ component of the Born charge
corresponding to the oxygen lattice. During the correlated calculations,
Wannier functions corresponding to $1s$ functions of both Li and
O were held frozen, while virtual excitations were carried out from
the $2s$ and $2p$ orbitals of oxygen anion. At the HF level we verified
that the symmetry relation $Z_{xx}^{*}=Z_{yy}^{*}=Z_{zz}^{*}$ holds,
along with the fact that the off-diagonal elements of the Born charge
tensor are negligible. Additionally, the optical sum rule $Z_{\alpha\beta}^{*}(\textrm{O})+2Z_{\alpha\beta}^{*}(\textrm{Li})=0$,
was also verified. 

\begin{figure}
\caption{One of the $2p$ valence Wannier functions of oxygen plotted along
the $(1,1,1)$ direction. In order to simulate the effect of optical
phonons, calculation was performed with oxygen shifted to the position
$(0.01a,0.01a,0.01a)$, and the Li positions unchanged. Nodes in the
Wannier function at positions $(\pm a/4,\pm a/4,\pm a/4)$ are due
to its orthogonality to the Li $1s$ Wannier functions located there. }
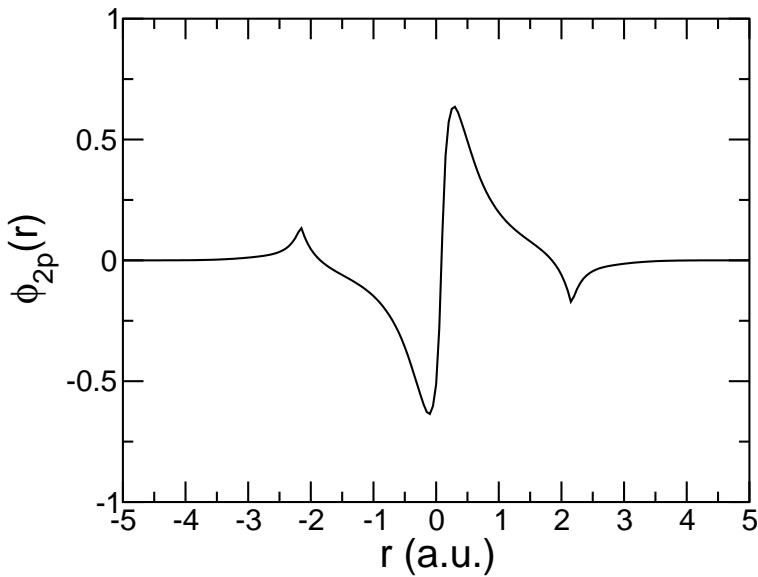
\vspace{1.5cm}

\includegraphics[width=10cm]{fig1}\label{fig-wan}
\end{figure}

Our calculated results for the Born charge of oxygen in Li$_{2}$O
are presented in table \ref{tab-born-li2o}. It is obvious from the
table that irrespective of the basis set used, most of the contribution
to the Born charge, as expected, is obtained at the HF level, with
relatively smaller contribution of electron-correlation effects. Correlation
effects do diminish the value of the Born charge as compared to its
HF value, however, rather insignificantly. Comparatively, speaking
the inclusion of one $d-$type basis function on oxygen has greater
influence on the Born charge of oxygen, and its value is increased
as compared to the one obtained using the $(s,p)$ basis set. With
the $(s,p,d)$ basis set, we found negligible influence on the Born
charge from the two-body contributions outside of the reference unit
cell. Thus our calculations, as per optical sum rule, predict a value
of Born charge of Li to be close to 0.91, in very good agreement with
the experimental results,\citep{osaka} with the error being less
than 5\%. In order to obtain a pictorial view of electronic structure
in the material, in Fig. \ref{fig-wan} we plot one of the oxygen
$2p$ valence Wannier functions, computed using the $(s,p,d)$ basis
set, along the $(1,1,1)$ direction. To simulate the effect of one
of the phonon modes, we displaced the oxygen atom of the primitive
cell to the position $(0.01a,0.01a,0.01a)$, where $a$ is the lattice
constant, leaving the Li atoms undisturbed. From the figure it is
obvious that the Wannier function is highly localized around the oxygen
site, with small value in the interatomic region, implying the ionic
character of the material.

\begin{table}
\caption{Correlated values of Born Charge of Li$_{2}$O obtained in our calculations.
Column with heading HF refers to results obtained at the Hartree-Fock
level. Heading {}``one-body'' refers to results obtained after including
the corrections due to {}``one-body'' excitations from each Wannier
function of the unit cell, to the HF value. Two-body (O) implies results
include additional corrections due to simultaneous excitations from
two distinct Wannier functions located on the anion in the reference
unit cell. Two-body (1NN) and two-body (2NN) correspond to two-body
correlation effects involving 1st and 2nd, nearest neighbor Wannier
functions, respectively. }

\begin{tabular}{|c|c|c|c|c|c|c|c|}
\hline 
Basis Set & \multicolumn{6}{c|}{Oxygen Born Charge} & Experiment\tabularnewline
\cline{3-3} \cline{4-4} \cline{5-5} \cline{6-6} \cline{7-7} \cline{8-8} 
\multicolumn{2}{|c|}{} & HF & One-body  & Two-Body (O) & Two-body (1NN) & Two-body (2NN) & 1.90$^{a}$\tabularnewline
\hline
\cline{3-3} \cline{4-4} \cline{5-5} \cline{6-6} \cline{7-7} \cline{8-8} 
\multicolumn{2}{|c|}{$(s,p)$} & 1.794 & 1.783 & 1.763 & 1.781 & 1.777 & \tabularnewline
\hline
\cline{3-3} \cline{4-4} \cline{5-5} \cline{6-6} \cline{7-7} \cline{8-8} 
\multicolumn{2}{|c|}{$(s,p,d)$} & 1.851 & 1.840 & 1.814 & --- & ---- & \tabularnewline
\hline
\end{tabular}\label{tab-born-li2o}

$^{a}$As deduced from the Szigeti charge reported in reference. \citep{osaka} 
\end{table}
 Another quantity which could help us determine whether the system
is ionic or covalent is the Mulliken population associated with the
atoms of the primitive cell. With the $(s,p)$ basis set, the Li and
O Mulliken populations were found to be $-2.005$ and $0.984$, respectively,
while with the $(s,p,d)$ basis set the populations were $-2.004$
and $0.984$. It is obvious that the computed Mulliken populations
have converged with respect to the basis set, and are very close to
the nominal ionicities of the two atoms. This further confirms the
ionic picture of the bonding in this substance.

\subsection{LiCl}

These calculations were performed using the basis set proposed by
Prencipe \emph{et al.}\citep{dovesi-licl} in their \emph{ab initio}
HF study of alkali halides. The experimental fcc geometry with lattice
constant of 5.07  \AA  \ was assumed. The unit cell employed during
the calculations consisted of anion placed at the origin, and the
cation at the position $(a/2,0,0)$. During the many-body calculations,
only the valence Wannier functions composed of $3s$, and $3p$ orbitals
localized on Cl, were correlated. In order to calculate the Born charge,
the anion was displaced in the $x$-direction by an amount $0.01a$.
For the finite-field computation of the dipole expectation value,
an external electric field of strength $\pm0.01$ a.u., also along
the $x$-axis, was applied. 

\begin{table}
\caption{Correlated values of Born Charge of LiCl obtained in our calculations.
Rest of the information is same as in the caption of table \ref{tab-born-li2o}.}

\begin{tabular}{|c|c|c|c|c|c|c|}
\hline 
Basis Set & \multicolumn{5}{c|}{Chlorine Born Charge} & Experiment\tabularnewline
\cline{3-3} \cline{4-4} \cline{5-5} \cline{6-6} \cline{7-7} 
\multicolumn{2}{|c|}{} & HF & One-body  & Two-Body (O) & Two-body (1NN) & 1.231$^{a}$\tabularnewline
\hline
\cline{3-3} \cline{4-4} \cline{5-5} \cline{6-6} \cline{7-7} 
\multicolumn{2}{|c|}{$(s,p)$} & 1.019 & 1.020 & 1.037 & 1.013 & \tabularnewline
\hline
\cline{3-3} \cline{4-4} \cline{5-5} \cline{6-6} \cline{7-7} 
\multicolumn{2}{|c|}{$(s,p,d)$} & 1.098 & 1.128 & 1.178 & --- & \tabularnewline
\hline
\end{tabular}

$^{a}$Computed from the experimental data reported in reference.
\citep{exp-licl}

\label{tab-born-licl}
\end{table}

The results of our calculations are presented in table \ref{tab-born-licl}.
With our $(s,p)$ basis set the correlation effects were included
up to the level of two-body increments involving the Wannier functions
in the nearest-neighboring cells. However, with the augmented $(s,p,d)$
basis set, inclusion of two-body increments beyond the reference unit
cell, did not make any significant difference to the results. Similar
to the case of, Li$_{2}$O, we note that: (a) inclusion of one $d$-type
function improves the results both at the HF and the correlated levels,
and (b) the inclusion of electron-correlation effects generally improves
the results as compared to the HF values. The final value of Born
charge of LiCl of 1.178 obtained from these calculations, is again
in very good agreement with the experimental value, with an error
smaller than 5\%.

\section{Conclusions}

\label{conclusion} In conclusion, we performed Wannier-function-based
\emph{ab initio} correlated calculations on crystalline Li$_{2}$O
and LiCl to compute their Born effective charges, using a methodology
recently developed in our group. Calculations were performed using
polarizable basis sets, and the results obtained were in very good
agreement with the experimental ones. Additionally, we also resolved
an ambiguity associated with rather small value of experimentally
measured effective charge of Li$_{2}$O, which had raised the possibility
of partial covalence in its chemical structure,\citep{osaka} by demonstrating
that the measured charge was the Szigeti charge, and not the Born
charge of the compound. The Born charge obtained from the Szigeti
charge was found to be in good agreement with its nominal ionic value.
By further computing the Mulliken charges and exploring the Wannier
functions, we demonstrated that the results were consistent with a
ionic picture of Li$_{2}$O. The computational effort involved in
our calculations is significant heavier than in the calculations performed
using the density-functional theory. This is because our approach
is many-body in nature, requiring calculations beyond the mean-field.
At present we are exploring the possibility of using our approach
to compute the Born charges of ferroelectric, as well as covalent
materials, and we will publish our results in future as and when they
become available.

\begin{acknowledgments}
This work was supported by  grant no. SP/S2/M-10/2000 from Department
of Science and Technology (DST), Government of India.
\end{acknowledgments}


\begin{thebibliography}{10}
\bibitem{dft-pol}See, \emph{e.g.}, R. Resta, Rev. Mod. Phys. \textbf{66},
899 (1994), and references therein.

\bibitem{king}R. D. King-Smith and D. Vanderbilt, Phys. Rev. B \textbf{47},
1651 (1993) .

\bibitem{resta}R. Resta, Europhys. Lett. \textbf{22}, 133 (1993).

\bibitem{priya-1}P. Sony and A. Shukla, Phys. Rev. B \textbf{70},
241103(R) (2004).

\bibitem{priya-2}P. Sony and A. Shukla, Phys. Rev. B \textbf{73},
165106 (2006).

\bibitem{hozoi}L. Hozoi, U. Birkenheuer, P. Fulde, A. Mitrushchenkov,
and H. Stoll, Phys. Rev. B \textbf{76}, 085109 (2007).

\bibitem{osaka}T. Osaka and I. Shindo, Solid State Commun. \textbf{51},
421 (1984).

\bibitem{farley}T. W. D. Farley, W. Hayes, S. Hull, M. T. Hutchings,
M. Alba, and M. Vrtis, Physica B \textbf{156} and \textbf{157}, 99
(1989).

\bibitem{shukla-born}A. Shukla, Phys. Rev. B \textbf{61}, 13277 (2000).

\bibitem{born}M. Born and K. Huang, \emph{Dynamical Theory of Crystal
Lattices} (Clarendon, Oxford, England, 1954).

\bibitem{mcweeny}See, \emph{e.g.}, R. McWeeny, \emph{Methods of Molecular
Quantum Mechanics}, 2nd edition (Academic Press, London, 1989).

\bibitem{shukla1}A. Shukla, M. Dolg, H. Stoll, and P. Fulde, Chem.
Phys. Lett. \textbf{262}, 213 (1996).

\bibitem{shukla2}A. Shukla, M. Dolg, P. Fulde, and H. Stoll, Phys.
Rev. B \textbf{57}, 1471 (1998).

\bibitem{shukla3}A. Shukla, M. Dolg, P. Fulde, and H. Stoll, J. Chem.
Phys. \textbf{108}, 8521 (1998).

\bibitem{shukla4}A. Shukla, M. Dolg, P. Fulde, and H. Stoll,  Phys.
Rev. B \textbf{60}, 5211 (1999). 

\bibitem{shukla5}A. Abdurahman, A. Shukla, and M. Dolg, J. Chem.
Phys. \textbf{112}, 4801 (2000). 

\bibitem{stoll}H. Stoll, Phys. Rev. B 46 (1992) 6700; H. Stoll, Chem.
Phys. Lett. \textbf{191}, 548 (1992).

\bibitem{szigeti}B. Szigeti, Trans. Faraday Soc. 45 (1949) 155; \emph{ibid.},\emph{
}Proc. R. Soc. A \textbf{204}, 51 (1950).

\bibitem{axe}J. D. Axe, Phys. Rev \textbf{139}, A 1215 (1965).

\bibitem{dovesi}R. Dovesi, C. Roetti, C.-Freyria-Fava, M. Prencipe,
and V. R. Saunders, Chem. Phys. \textbf{156}, 11 (1991).

\bibitem{dovesi-licl}M. Prencipe, A. Zupan, R. Dovesi, E. Apr\'a,
and V. R. Saunders, Phys. Rev. B \textbf{51}, 3391 (1995).

\bibitem{exp-licl}M. J. L. Sangster, U. Schr\"oder, and R. M. Atwood,
J. Phys. C \textbf{11}, 1523 (1978).
\end{thebibliography}
\end{document}